\documentclass[twocolumn,secnumarabic,amssymb, nobibnotes, aps, prd]{revtex4-2}

\newcommand{\mycomment}[1]{}
\newtheorem{theorem}{Theorem}[section]
\newtheorem{lemma}{Lemma}[section]
\usepackage{amsmath, braket, graphicx, xcolor, ulem}
\graphicspath{{./images/}}

\begin{document}

\title{Gaussian Quantum Illumination via Monotone Metrics}%

\author{Dong Hwan Kim}%
\thanks{These authors contributed equally.}
\author{Yonggi Jo}
\thanks{These authors contributed equally.}
\author{Duk Y. Kim}
\author{Taek Jeong}
\author{Jihwan Kim}
\author{Nam Hun Park}
\author{Zaeill Kim}
\author{Su-Yong Lee}
\email[Su-Yong Lee: ]{suyong2@add.re.kr}
\affiliation{Agency for Defense Development, Daejeon 34186, Korea}

\begin{abstract}
Quantum illumination is to discern the presence or absence of a low reflectivity target, where the error probability decays exponentially in the number of copies used.
When the target reflectivity is small so that it is hard to distinguish target presence or absence, the exponential decay constant falls into a class of objects called monotone metrics.
We evaluate monotone metrics restricted to Gaussian states in terms of first-order moments and covariance matrix.
Under the assumption of a low reflectivity target, we explicitly derive analytic formulae for decay constant of an arbitrary Gaussian input state.
Especially, in the limit of large background noise and low reflectivity, there is no need of symplectic diagonalization which usually complicates the computation of decay constants.
First, we show that two-mode squeezed vacuum (TMSV) states are the optimal probe among pure Gaussian states with fixed signal mean photon number.
Second, as an alternative to preparing TMSV states with high mean photon number, we show that preparing a TMSV state with low mean photon number and displacing the signal mode is a more experimentally feasible setup without degrading the performance that much.
Third, we show that it is of utmost importance to prepare an efficient idler memory to beat coherent states and provide analytic bounds on the idler memory transmittivity in terms of signal power, background noise, and idler memory noise.
Finally, we identify the region of physically possible correlations between the signal and idler modes that can beat coherent states.
\end{abstract}
\maketitle

\section{Introduction}
Quantum illumination (QI) seeks to discern the presence or absence of a low reflectivity target by utilizing an additional idler mode of light to enhance the detection probability \cite{SL08, ST08, SG09, EL13, QZ17, AK20, SB20, HY21, SL22, SZ14, LF18, GP18, RN20, MB21}.
Coherent states are the benchmark to compare the performance of QI to classical illumination, where one uses only one mode of light to detect a target.
It was first shown in  Ref. \cite{ST08} that two-mode squeezed vacuum (TMSV) states can outperform coherent states.
The performance is quantified by the ability to distinguish two different quantum states $\rho_0$ and $\rho_1$ each corresponding to the state of target absent and present.

The Helstrom bound gives the minimum total uncertainty when distinguishing $\rho_0$ and $\rho_1$ \cite{CH69}.
The quantum Chernoff bound indicates the ability to distinguish $\rho_0$ and $\rho_1$ in the asymptotic limit of symmetric hypothesis testing, when we have $N$ copies of the same state and optimal collective measurement on all $N$ copies is possible.
More precisely, it states that the Helstrom bound of the tensor product states $\rho_{0,1}^{\otimes N}$ decays exponentially in $N$  \cite{KA07}:
\begin{align}
\gamma_{\text{col}} &= \lim_{N\rightarrow \infty} - \frac{1}{N} \text{Log } P_{err} \left( \rho_0^{\otimes N}, \rho_1^{\otimes, N} \right)\\
&= -\min_{0\leq s\leq 1} \text{Log Tr } \rho_0^s \rho_1^{1-s},  \label{QCB:def}
\end{align}
where $P_{err}$ denotes the Helstrom bound.
We refer to $\gamma_{\text{col}}$ as the collective decay constant.
TMSV states can show a maximum of 6 dB enhancement in this decay constant compared to a coherent state with same signal power in the regime of low signal power, low target reflectivity and high background noise \cite{ST08}.

As performing collective measurements on all copies at once is typically hard, one would seek the behavior of error probabilities when only local measurements combining each incoming signal and idler mode are possible.
Given a local measurement $\hat{O}$, we have two probability distributions $\mu_0$ and $\mu_1$ corresponding to $\rho_0$ and $\rho_1$, and the problem is to distinguish between the two probability distributions given $N$ samples drawn from either distribution.
The error probability again decays exponentially, and when the two states are close, the decay constant is given by the signal-to-noise ratio $(m_0-m_1)^2/(2(\sigma_0 +\sigma_1)^2)$ \cite{HC52}.
Here, $m_i$ and $\sigma_i^2$ are the mean and variance of the probability distribution $\mu_i$ $(i=0,1)$.
We define the local decay constant $\gamma_{\text{loc}}$ to be the decay constant maximized over all possible local measurements $\hat{O}$.
However, no simple formula like Eq. (\ref{QCB:def}) seems to be known for $\gamma_{\text{loc}}$ in general.

Computation of the decay constants is complicated in general, even when considering only Gaussian states \cite{JC08, PM16, KS18}.
Gaussian states, which include coherent states, TMSV states, and thermal states, have been considered in various works of QI for their ease in both experimental and theoretical aspects \cite{ST08, SG09, EL13, QZ17, AK20, SB20, HY21, SL22}.
For Gaussian states, the main obstacle is the necessity of symplectic diagonalization of the two output states \cite{SP08, ST08, AK20} or computation of the symmetric logarithmic derivative \cite{MP09, DS19, RJ22}.
These procedures make analytic formulae of decay constants highly complicated, if exists.

In QI, the low reflectivity condition implies that the difference between the two states of interest is infinitesimal.
In this case, the decay constants also become infinitesimal quantities and give infinitesimal distances between quantum states \cite{KA07, JC08}.
These distances are examples of monotone metrics, infinitesimal distances between quantum states which decrease under completely positive trace preserving (CPTP) maps \cite{DP96}.
Taking advantage of this, we compute $\gamma_{\text{col}}$ and $\gamma_{\text{loc}}$ directly in terms of first-order moments and covariance matrix of the input two-mode Gaussian state.
Our results only need to compute the symplectic diagonalization of the idler mode, which are in contrast to previous methods requiring symplectic diagonalization of the two-mode Gaussian states $\rho_0$ and $\rho_1$.
We also show that in the large background noise limit, no symplectic diagonalization is needed to compute the decay constants.
We apply our results to analyze the effect of single-mode Gaussian unitaries on both signal and idler modes on the performance of QI.

This paper is organized as follows.
In Sec. \ref{Sec:GS}, we provide a short review on Gaussian states and quantum illumination.
In Sec. \ref{Sec:PM}, the results of Petz on monotone metrics are summarized and its restriction to Gaussian states is computed in terms of first-order moments and covariance matrix. 
As $\gamma_{\text{col}}$ and $\gamma_{\text{loc}}$ are described by monotone metrics, we provide in Sec. \ref{Sec:DC} analytic formulae for the decay constants in terms of first-order moments and covariance matrix.
These are used to show that TMSV states are indeed optimal for QI among pure Gaussian states with same signal power.
Also we investigate the effect of single-mode Gaussian operations on the signal mode to enhance QI performance.
In Sec. \ref{Sec:Loss}, we consider the effect of idler loss on the performance of QI.
We show that single-mode Gaussian operations on the idler mode are not advantageous to overcome idler loss and identify the region of idler loss and noise where observing quantum advantage over coherent states is possible.
In Sec. \ref{Sec:Corr}, we identify the region of correlations between the signal and idler modes of QI needed to see quantum advantage.
Sec. \ref{Sec:Con} gives a summary of the results.

\section{Gaussian Quantum Illumination} \label{Sec:GS}
\subsection{Review of Gaussian States}
Gaussian states are determined solely by their first- and second-order moments.
To describe an $n$-mode Gaussian state, we write the canonical operators as $\hat{r} = (\hat{x}_1, \hat{p}_1, \cdots, \hat{x}_n, \hat{p}_n)^T$.
Then the commutation relation is
\begin{equation}
[\hat{r}, \hat{r}^T] = i \Omega = i \oplus^n \begin{pmatrix} 0 & 1 \\ -1 & 0 \end{pmatrix}.
\end{equation}
The first-order moments $\overline{r} = \braket{\hat{r}}$ and the covariance matrix $V= \braket{\{ \hat{r} - \overline{r}, \hat{r}^T - \overline{r}^T\}}$ fully determine a Gaussian state.
The only constraints on the first-order moments and covariance matrix are the uncertainty relation $V+i\Omega \geq 0$ and symmetry of the covariance matrix.
For example, TMSV states are two-mode Gaussian states with zero first-order moments and covariance matrix
\begin{equation} \label{covmat:TMSV}
V=\begin{pmatrix}
1+2N_S & 2\sqrt{N_S+N_S^2} \sigma_z \\
2\sqrt{N_S+N_S^2} \sigma_z & 1+2N_S
\end{pmatrix},
\end{equation}
where $\sigma_z$ is the Pauli $z$ matrix.

Given a symmetric $2n\times 2n$ positive matrix $V$, there is a matrix $S$ and numbers $\nu_i \geq 0\;(i=1,\cdots,n)$ such that
\begin{align}
V &= S\text{ diag} (\nu_1,\nu_1, \nu_2, \nu_2, \cdots, \nu_n,\nu_n) S^T,\\
\Omega &= S\Omega S^T. \label{def:SP}
\end{align}
The uncertainty relation is equivalent to that the symplectic eigenvalues $\nu_i$ satisfy $\nu_i \geq 1$. Eq. (\ref{def:SP}) means that $S$ is an element of the symplectic group $Sp(n)$.
For each $S\in Sp(n)$, there is a unitary operator $\hat{S}$ (defined up to sign) whose action on canonical operators is $\hat{S} \hat{r} \hat{S}^\dagger = S \hat{r}$.
The operators $\hat{S}$ include local phase shifts, squeezing, and beam splitting operations.
For a $2n$ column vector $\overline{r}$, one can define the displacement operator $\hat{D}_{\overline{r}} = \text{exp}(i \overline{r}^T \Omega \hat{r})$ so that $\hat{D}_{\overline{r}} \hat{r} \hat{D}_{\overline{r}}^\dagger = \hat{r} + \overline{r}$.
The action of both $\hat{S}$ and $\hat{D}$ preserves Gaussian states.
If $\rho$ is a Gaussian state with first-order moments $\overline{r}_1$ and covariance matrix $V$, then $\hat{D}_{\overline{r}_2}^\dagger \hat{S}^\dagger \rho \hat{S} \hat{D}_{\overline{r}_2}$ has first-order moments $\overline{r}_2 + S \overline{r}_1$ and covariance matrix $SVS^T$.
Therefore, any Gaussian state can be constructed from a thermal state and applying suitable $\hat{S}$ and $\hat{D}$ sequentially.
More details on Gaussian states and the symplectic diagonalization can be found in Ref. \cite{AS17}.

$n$-mode Gaussian states form a $n(2n+3)$ dimensional space.
The $n(2n+3)$ degrees of freedom of Gaussian states can be attributed to $n$ from the symplectic eigenvalues $\nu_i$, $2n$ from single-mode squeezing on each mode, $n(n-1)$ from two-mode squeezing on each pair of modes, $n(n-1)$ from beam splitting on each pair of modes, and $2n$ from displacement on each mode.
Each of these degrees of freedom gives rise to infinitesimal directions in the space of Gaussian states, and we will see in Sec. \ref{Sec:PM} that each of these directions contributes independently to the decay constants.

\subsection{Gaussian States in Quantum Illumination} \label{SSec:GQI}
\begin{figure}[t]
\includegraphics[width=0.49\textwidth]{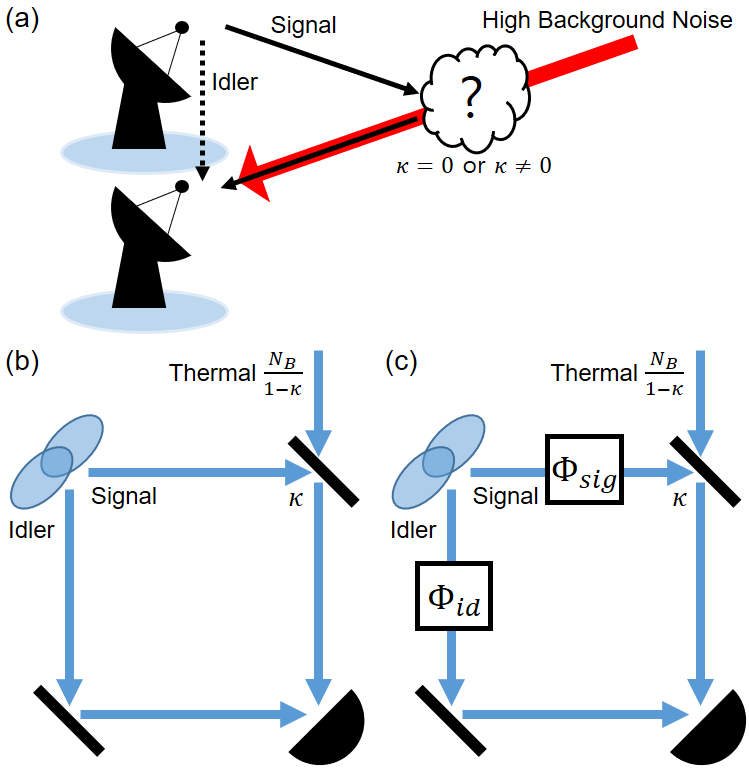}
\caption{(a) A general depiction of QI.
The signal mode of a two-mode state is sent to interrogate a region with high background noise where a low-reflectivity target might exist.
The reflected signal is measured with the idler to determine whether the target is present or not.
(b) The high background noise is modelled as a thermal state with mean photon number $N_B/(1-\kappa)$ and the low-reflectivity target is modelled as a beam splitter with reflectivity $\kappa\ll1$.
(c) We generalize the model by applying single-mode Gaussian channels on both modes and compute the decay constants.
For the signal mode, we consider unitary operations which increase the mean photon number as an alternative to preparing TMSV states with high signal mean photon number.
For the idler mode, we consider mixing with a thermal state through a beam splitter to model possibly existent loss or noise in the idler memory.
}
\label{Fig:QI}
\end{figure}

Consider the detection of a low reflectivity target under high background noise as shown in Fig. \ref{Fig:QI} (a).
We model the low reflectivity target as a beam splitter of reflectivity $\kappa$ as shown in Fig. \ref{Fig:QI} (b).
One mode of a two-mode source is sent to the target, where a small portion of the signal mode is reflected and mixed with a thermal background.
The thermal background is taken to be $\kappa$ dependent so that the thermal noise the detector observes is $\kappa$ independent.
If the input two-mode state is Gaussian, then the two output states, depending on target absent or present, are also Gaussian.
Therefore, we refer to QI using a Gaussian input as Gaussian QI (GQI).

Suppose a general two-mode Gaussian state with first-order moments $\overline{r}=(x_s,p_s,x_i,p_i)^T$ and covariance matrix $V=(a_{ij})$ is used in GQI.
Under the presence of a target with low reflectivity $\kappa$, the returned state is also a Gaussian state described by
\begin{align} \begin{split}
\overline{r}_r (\kappa) &= X \overline{r},\hspace{2mm} V_r (\kappa) = X V X^T + Y,\\
X&=\text{diag} (\sqrt{\kappa},\sqrt{\kappa},1,1),\\
Y&=\text{diag}(1-\kappa+2N_B,1-\kappa+2N_B,0,0).
\end{split} \end{align}
The two output states we wish to distinguish correspond to the cases of $\kappa=0$ ($\rho_0$, target absent) and $0\neq\kappa \ll1$ ($\rho_1$, target present).
Since we are assuming $\kappa\ll1$, the leading order differences in $\overline{r}_r$ and $V_r$ between $\kappa =0$ and non-zero $\kappa$ are
\begin{align} \begin{split} \label{dQI}
\sqrt{\kappa} d\overline{r} &= \sqrt{\kappa} (x_s,p_s,0,0)^T,\\
\sqrt{\kappa} dV &= \sqrt{\kappa} \begin{pmatrix} & &a_{13}& a_{14} \\  &  & a_{23} & a_{24} \\ a_{13} & a_{23} &  &  \\ a_{14}& a_{24} &  &  \end{pmatrix}.
\end{split} \end{align}
Here and henceforth, blanks in a matrix mean that its entry is zero.
The symplectic eigenvalues of $V_r(0)$ are $\nu_1 = 1+2N_B$ and $\nu_2^2 = a_{33}a_{44}-a_{34}^2$. Analytic formula of symplectic diagonalization of $V_r(0)$ in terms of $a_{33}, a_{44}, a_{34}$ exist, but is too lengthy to explicitly write down. These will be the input data to the results of Sec. \ref{Sec:PM} to compute the decay constants $\gamma_{\text{col}}$ and $\gamma_{\text{loc}}$.

As shown in Fig. \ref{Fig:QI} (b), QI assumes that the idler mode is preserved ideally to be measured with the returned signal mode.
We extend the QI model by applying single-mode Gaussian channels $\Phi_{sig(id)}$ on the signal (idler) mode, as in Fig. \ref{Fig:QI} (c).
In Sec. \ref{Sec:DC}, we study the effect of single mode unitaries on the signal mode which increase the mean photon number in hope to increase the decay constants.
In Sec. \ref{Sec:Loss}, we study the effect of loss or noise which will be in a realistic idler channel.

\section{Monotone Metrics on Gaussian States} \label{Sec:PM}
Infinitesimal distances on a space are described by a metric, which is an inner product on the tangent space at each point of the space.
We are interested in the space of invertible density matrices or the space of invertible Gaussian states.
For the former case, the tangent space at each point is the collection of Hermitian traceless matrices.
We denote the inner product as $g_{\rho}$ for each invertible density matrix $\rho$.
If the metric decreases under all completely positive trace preserving (CPTP) maps, we call the metric to be monotone:
\begin{equation}
g_{\Phi(\rho)} (\Phi(A), \Phi(A)) \leq g_\rho (A,A)
\end{equation}
where $\Phi$ is a CPTP map and $A$ is a tangent vector at $\rho$.
When $\Phi$ is a unitary evolution, hence invertible, applying the monotone property twice with $\Phi$ and $\Phi^{-1}$ implies that the inner product is invariant under unitaries:
\begin{equation}
g_{\Phi(\rho)} (\Phi(A), \Phi(A)) = g_\rho (A,A).
\end{equation}
Monotone metrics arise naturally from general quantum divergences.
Let $D(\rho||\sigma)$ be a quantum divergence monotone under CPTP maps, for example, the relative entropy or R\' enyi entropies.
Then the divergence between states with infinitesimal difference gives rise to a monotone metric as $D(\rho||\rho +\epsilon\;d\rho) = \epsilon^2 g_\rho (d\rho, d\rho) + o(\epsilon^2)$ \cite{AL99}.

Petz classified monotone metrics on finite dimensional quantum spaces \cite{DP96}.
For each monotone metric $g$, there is an operator monotone function $f$ satisfying $f(t) = tf(t^{-1})$ such that $g_\rho$ is given by
\begin{align}
g_\rho (A,B) &= \text{Tr } A^\dagger c_f (\mathbf{L}_\rho, \mathbf{R}_\rho)(B), \\
c_f(x,y) &= 1/(y f(x/y)),
\end{align}
where $A, B$ are tangent vectors at $\rho$.
The definition of an operator monotone function, which is immaterial in this paper, can be found in Ref. \cite{FK80}.
Here, $\mathbf{L}_\rho$ and $\mathbf{R}_\rho$ are left and right multiplication by $\rho$ respectively.
If $\rho$ is diagonalized as $\rho = \sum \lambda_n \ket{\psi_n}\bra{\psi_n}$ with $\braket{\psi_n|\psi_m} = \delta_{nm}$, then the metric is
\begin{equation}
g_\rho (A,B) = \sum_{n,m} c_f (\lambda_n, \lambda_m)  A_{nm}^* B_{nm}
\end{equation}
where $A_{nm} = \braket{\psi_n | A | \psi_m}$ are matrix elements of $A$ and $B_{nm}$ are analogously defined.
Infinitesimal values of the collective and local decay constants $\gamma_{\text{col(loc)}}$ are related to monotone metrics and their associated functions are given by $f_{\text{col}} = 2(\sqrt{t}+1)^2$ and $f_{\text{loc}} = 4(t+1)$ respectively \cite{JC08, AL99}.

In the following, we compute the monotone metric associated to an operator monotone function $f$ restricted to Gaussian states.
We first determine the metric in terms of infinitesimal changes in symplectic eigenvalues and infinitesimal unitaries, and then write the expression in terms of first-order moments and covariance matrix.
From unitary invariance, it suffices to compute the metric at diagonal states, which are thermal states.
Let $\rho$ be an $n$-mode thermal state with symplectic eigenvalues $\nu_1, \cdots, \nu_n>1$.
The tangent vectors at $\rho$ can be classified into two types, ones which commute with $\rho$ and ones which do not.
The tangent vectors which commute with $\rho$ arise from infinitesimal changes in $\nu_i$:
\begin{equation}
\frac{\partial \rho}{\partial \nu_i} = \left( \frac{2\hat{a}_i^\dagger \hat{a}_i}{\nu_i^2-1}  -\frac{1}{\nu_i+1} \right) \rho.
\end{equation}
The tangent vectors which do not commute with $\rho$ arise from infinitesimal unitaries:
\begin{equation}
i[\hat{H}, \rho],
\end{equation}
where $\hat{H}$ is a Hamiltonian generator describing displacement, squeezing or beam splitting action.

Unitary invariance combined with the polarization identity is a powerful tool to show certain tangent vectors are orthogonal to each other.
Given a symmetric real bilinear form $g$, the polarization identity implies
\begin{equation}
g(u+v,u+v) = g(u-v, u-v) \Rightarrow g(u,v)=0
\end{equation}
for vectors $u,v$.
Thermal states and tangent vectors $\tfrac{\partial\rho}{\partial \nu_i}$ are invariant under local phase shifts, while the tangent vectors $i[\hat{H}, \rho]$ are not.
This immediately gives
\begin{equation}
g_\rho \left( \frac{\partial\rho}{\partial\nu_i}, i [\hat{H}, \rho] \right) =0.
\end{equation}
Furthermore, the $2n(n+1)$ tangent vectors $i[\hat{H}, \rho]$ with
\begin{align}
\hat{H} &= \hat{x}_i,\; \hat{p}_i,\;
\frac{1}{2} \left(  \hat{a}_i^2 + \hat{a}_i^{\dagger 2} \right),\;
\frac{i}{2} \left( \hat{a}_i^2 -  \hat{a}_i^{\dagger 2} \right), \;
\hat{a}_i \hat{a}_j + \hat{a}_i^{\dagger} \hat{a}_j^\dagger,\; \nonumber \\
& i \left(  \hat{a}_i \hat{a}_j - \hat{a}_i^{\dagger} \hat{a}_j^\dagger \right),\;
\hat{a}_i^{\dagger} \hat{a}_j + \hat{a}_i \hat{a}_j^\dagger ,\;
i \left( \hat{a}_i^{\dagger} \hat{a}_j -  \hat{a}_i \hat{a}_j^\dagger \right),  \label{ONB}
 \end{align}
are all mutually orthogonal. For example, a local $\pi/2$ phase shift on mode $i$ maps $\hat{x}_i \mapsto \hat{p}_i$, $\hat{p}_i \mapsto -\hat{x}_i$, so unitary invariance and the polarization identity imply
\begin{align} \begin{split}
g_{\rho} &(i[\hat{x}_i+\hat{p}_i, \rho], i[\hat{x}_i+\hat{p}_i, \rho])\\
& \hspace{20mm} = g_\rho (i[\hat{p}_i - \hat{x}_i, \rho],i[\hat{p}_i - \hat{x}_i, \rho]) \\
&\Rightarrow \; g_\rho  (i[\hat{x}_i, \rho], i[\hat{p}_i, \rho])=0.
\end{split} \end{align}
Therefore, it suffices to determine the metric on tangent vectors $\tfrac{\partial\rho}{\partial\nu_i}$ and the values of $g_\rho (i[\hat{H}, \rho], i[\hat{H}, \rho])$ for each $\hat{H}$ from Eq. (\ref{ONB}).

We determine the metric on tangent vectors $\tfrac{\partial\rho}{\partial\nu_i}$. As the tangent vectors commute with $\rho$, we have
\begin{align}
c_f (\mathbf{L}_\rho, \mathbf{R}_\rho) \left(\frac{\partial \rho}{\partial \nu_i} \right) &=\frac{1}{f(1)} \left(\frac{2\hat{a}_i^\dagger \hat{a}_i}{\nu_i^2-1}  -\frac{1}{\nu_i+1} \right). 
\end{align}
This leads to
\begin{align}
g_\rho \left( \frac{\partial \rho}{\partial \nu_i}, \frac{\partial \rho}{\partial \nu_i}\right) &= \frac{1}{f(1)}\frac{1}{\nu_i^2-1}, \label{norm_ev:nu} \\
g_\rho \left( \frac{\partial \rho}{\partial \nu_j}, \frac{\partial \rho}{\partial \nu_i}\right) &=0 \hspace{5mm}(i\neq j).
\end{align}
Therefore, the metric is also diagonal in terms of infinitesimal changes of $\nu_i$ and we have diagonalized the metric.
Note that Eq. (\ref{norm_ev:nu}) only depends on $f(1)$, not the whole shape of $f$.
This is reminiscent of the up-to-scale uniqueness of monotone metrics on classical probabilities \cite{LC86, DP96}.

To determine $g_\rho (i[\hat{H}, \rho], i[\hat{H}, \rho])$, first observe that unitary invariance from local phase shifts implies
\begin{equation}
g_\rho (i[\hat{x}_i, \rho], i[\hat{x}_i, \rho])=g_\rho (i[\hat{p}_i, \rho], i[\hat{p}_i, \rho])
\end{equation}
and analogous equalities hold for phase conjugate Hamiltonians in Eq. (\ref{ONB}).
We may also trace out all but one or two systems, as a multi-mode thermal state is a tensor product of one-mode thermal states.
This means that it suffices to compute the following four quantities:
\begin{align} \begin{split}
&g_\rho (i[\hat{x}_1, \rho], i[\hat{x}_1, \rho]), \label{norm:unitary}\\
& g_\rho \left(i \left[ \frac{1}{2} \left( \hat{a}_1^2+\hat{a}_1^{\dagger 2} \right), \rho \right], i\left[\frac{1}{2}  \left( \hat{a}_1^2+\hat{a}_1^{\dagger 2} \right), \rho \right] \right), \\
& g_\rho \left(i \left[\hat{a}_1 \hat{a}_2 +\hat{a}_1^{\dagger} \hat{a}_2^\dagger, \rho \right], i\left[\hat{a}_1 \hat{a}_2 + \hat{a}_1^{\dagger} \hat{a}_2^\dagger, \rho \right] \right),\\
& g_\rho \left(i \left[\hat{a}_1^{\dagger} \hat{a}_2 + \hat{a}_1 \hat{a}_2^\dagger , \rho \right], i\left[\hat{a}_1^{\dagger} \hat{a}_2 + \hat{a}_1 \hat{a}_2^\dagger , \rho \right] \right),
\end{split} \end{align}
where $\rho$ is a one-mode thermal state with symplectic eigenvalue $\nu$ or a two-mode thermal state with symplectic eigenvalues $\nu_1$ and $\nu_2$.

As a computational tool, we use the Bargmann representation, which converts a density matrix into a function of complex variables.
Essential properties of this representation are summarized in Appendix \ref{App:BR}.
A one-mode thermal state with covariance matrix $\text{diag}(\nu, \nu)$ in number basis is given as
\begin{equation}
\rho = \frac{2}{\nu+1} \sum_{n=0}^\infty \left( \frac{\nu-1}{\nu+1}\right)^n \ket{n}\bra{n},
\end{equation}
so the Bargmann representation is
\begin{equation}
K(z, w^*) = \frac{2}{\nu+1} \text{Exp} \left( \frac{\nu-1}{\nu+1} z w^* \right).
\end{equation}
Since thermal states are diagonal in number basis, the action of $c_f(\mathbf{L}_\rho, \mathbf{R}_\rho)$ is diagonal on $\ket{n}\bra{m}$:
\begin{align} \begin{split}
&c_f(\mathbf{L}_\rho, \mathbf{R}_\rho)(\ket{n}\bra{m})\\
&=\frac{\nu+1}{2} \left( \frac{\nu+1}{\nu-1} \right)^m f\left( \left( \frac{\nu-1}{\nu+1} \right)^{n-m} \right)^{-1} \ket{n}\bra{m}.
\end{split} \end{align}
Therefore, we can obtain the Bargmann representation of both $i[\hat{x}, \rho]$ and $c_f(\mathbf{L}_\rho, \mathbf{R}_\rho) (i[\hat{x}, \rho])$.
We have
\begin{align} \begin{split}
&i[\hat{x}, \rho] \\
&\hspace{2mm} \iff \frac{2\sqrt{2}i(z-w^*)}{(\nu+1)^2} \text{Exp} \left( \frac{\nu-1}{\nu+1}z w^* \right),
\end{split}\\ \begin{split} \label{Br:x}
&c_f(\mathbf{L}_\rho, \mathbf{R}_\rho) (i[\hat{x}, \rho])\\
&\hspace{2mm} \iff \frac{\sqrt{2}i(z-w^*)}{\nu-1}f\left( \frac{\nu+1}{\nu-1} \right)^{-1} \text{Exp} \left( z w^* \right).
\end{split}
\end{align}
The symmetry of $f$, $f(t) = tf(t^{-1})$, is used in obtaining Eq. (\ref{Br:x}).
\begin{widetext}
Now, computing $g_\rho (i[\hat{x}, \rho], i[\hat{x}, \rho])$ surmounts to evaluating a single integral:
\begin{align} \begin{split}
g_\rho (i[\hat{x}, \rho], i[\rho, \hat{x}]) &=  \int \frac{d^2 z\, d^2w}{\pi^2} \frac{4(z-w^*)(z^*-w)}{(\nu+1)^2(\nu-1)} f \left( \frac{\nu+1}{\nu-1} \right)^{-1} \text{Exp} \left(\frac{\nu-1}{\nu+1}zw^* + z^* w - |z|^2-|w|^2 \right)\\
&= \frac{2}{\nu-1} f \left( \frac{\nu+1}{\nu-1} \right)^{-1}. \label{norm_ev:x}
\end{split} \end{align}
The other values appearing in Eq. (\ref{norm:unitary}) are calculated in a similar procedure.
The following general integral formula is useful:
\begin{align} \begin{split}
&\int \frac{d^2z\, d^2w}{\pi^2} z^n z^{*m} w^k w^{*\ell} \text{Exp} \left( \alpha z w^* + \beta z^* w - |z|^2  - |w|^2 \right)\\
&\hspace{5mm}= \delta_{n+k,m+\ell}\; \alpha^{\text{max}(m-n,0)} \beta^{\text{max}(n-m,0)} \frac{a!\; b!}{|n-m|!} \,_2 F_1 \left( a+1,b+1;\; |n-m|+1;\; \alpha\beta  \right). \hspace{5mm} (\text{Re}(\alpha \beta)<1)
\end{split} \end{align}
Here, $a= \text{max}(n,m)$, $b=\text{max}(k,\ell)$, and $_2 F_1$ is the hypergeometric function.
The results are
\begin{align}
g_\rho \left(i \left[\frac{1}{2} \left( \hat{a}^2+\hat{a}^{\dagger 2} \right), \rho \right], i\left[\frac{1}{2}  \left( \hat{a}^2+\hat{a}^{\dagger 2} \right), \rho \right] \right)&=\frac{4\nu^2}{(\nu-1)^2} f\left( \frac{(\nu+1)^2}{(\nu-1)^2} \right)^{-1} ,\label{norm_ev:sms}\\
g_\rho \left(i \left[\hat{a}_1 \hat{a}_2 +\hat{a}_1^{\dagger} \hat{a}_2^\dagger, \rho \right], i\left[\hat{a}_1 \hat{a}_2 + \hat{a}_1^{\dagger} \hat{a}_2^\dagger, \rho \right] \right)&= \frac{2(\nu_1+\nu_2)^2}{(\nu_1-1)(\nu_2-1)} f\left(\frac{\nu_1+1}{\nu_1-1} \frac{\nu_2+1}{\nu_2-1} \right)^{-1},\label{norm_ev:tms}\\
g_\rho \left(i \left[\hat{a}_1^{\dagger} \hat{a}_2 + \hat{a}_1 \hat{a}_2^\dagger  , \rho \right], i\left[\hat{a}_1^{\dagger} \hat{a}_2 + \hat{a}_1 \hat{a}_2^\dagger , \rho \right] \right)&=  \frac{2(\nu_1-\nu_2)^2}{(\nu_1+1)(\nu_2-1)} f\left(\frac{\nu_1-1}{\nu_1+1} \frac{\nu_2+1}{\nu_2-1} \right)^{-1} .\label{norm_ev:tbs}
\end{align}
In Eq. (\ref{norm_ev:sms}), $\rho$ is a one-mode thermal state with covariance matrix $\text{diag}(\nu, \nu)$, while in Eqs. (\ref{norm_ev:tms}, \ref{norm_ev:tbs}), $\rho$ is a two-mode thermal state with covariance matrix $\text{diag}(\nu_1, \nu_1, \nu_2, \nu_2)$.
Eqs. (\ref{norm_ev:nu}, \ref{norm_ev:x}, \ref{norm_ev:sms}--\ref{norm_ev:tbs}) fully determine the monotone metric associated to an operator monotone function $f$ restricted to Gaussian states.
\end{widetext}

We have computed the metric in terms of infinitesimal changes in symplectic eigenvalues and infinitesimal unitaries.
We now write the results in terms of infinitesimal changes in first-order moments and the covariance matrix.
In terms of first-order moments and covariance matrix, a tangent vector is represented by a $2n$ column vector $d\overline{r}$ and a symmetric $2n\times 2n $ real matrix $dV$.
Let $E_{i,j}$ be the $2n\times 2n$ matrix where the $(i,j)$ entry is 1 and all other entries are 0.
Define the following matrices:
\begin{align}\begin{split}   \label{mat_basis}
N_i &:= E_{2i-1,2i-1} + E_{2i,2i}, \\
S_i &:= E_{2i,2i-1} + E_{2i-1,2i},\\
T_i &:= E_{2i-1,2i-1} - E_{2i,2i},\\
S_{ij} &:=E_{2j,2i-1} + E_{2j-1,2i} +E_{2i,2j-1}+ E_{2i-1,2j},\\
T_{ij} &:=E_{2j-1,2i-1}-E_{2j,2i}+E_{2i-1,2j-1}-E_{2i,2j},\\
A_{ij} &:= E_{2j,2i-1} - E_{2j-1,2i} -E_{2i,2j-1}+ E_{2i-1,2j},\\
B_{ij} &:= E_{2j-1,2i-1}+E_{2j,2i}+E_{2i-1,2j-1}+E_{2i,2j}.
\end{split}\end{align}
The indices run over $1\leq i \leq n$ or $1\leq i \neq j \leq n$.

Given a thermal state $\rho$ and its covariance matrix $V$, we have $\tfrac{\partial V}{\partial \nu_i} = N_i$, so $\tfrac{\partial \rho}{\partial \nu_i}$ corresponds to $N_i$.
Under unitary transformations $\rho \mapsto e^{i\hat{H} t} \rho e^{-i\hat{H} t}$ where $\hat{H}$ is a homogeneous quadratic operator, the covariance matrix transforms as $V\mapsto e^{X t} V e^{X^T t}$ where $X$ is defined by $-i[\hat{H}, \hat{r}_i] = \sum_{j=1}^{2n}X_{ij} \hat{r}_j$, for $1\leq i \leq 2n$.
In terms of tangent vectors, $i[\hat{H}, \rho]$ corresponds to $XV+VX^T$. Using that $V$ is diagonal, we have the following correspondences:
\begin{align} \begin{split}
i\left[ \frac{1}{2} \left(\hat{a}_i^2 +  \hat{a}_i^{\dagger 2}  \right), \rho \right] &\iff 2\nu_i S_i,\\
i\left[\frac{i}{2} \left( \hat{a}_i^2 -  \hat{a}_i^{\dagger 2} \right), \rho \right] &\iff 2\nu_i T_i,\\
i\left[\hat{a}_i \hat{a}_j + \hat{a}_i^{\dagger} \hat{a}_j^\dagger, \rho \right] &\iff (\nu_i+\nu_j) S_{ij},\\
i\left[i \left( \hat{a}_i \hat{a}_j - \hat{a}_i^{\dagger} \hat{a}_j^\dagger  \right), \rho \right] &\iff (\nu_i+\nu_j) T_{ij},\\
i\left[\hat{a}_i^{\dagger} \hat{a}_j + \hat{a}_i \hat{a}_j^\dagger, \rho \right] &\iff (\nu_i-\nu_j) A_{ij},\\
i\left[i \left( \hat{a}_i^{\dagger} \hat{a}_j -  \hat{a}_i \hat{a}_j^\dagger \right), \rho \right] &\iff (\nu_i-\nu_j) B_{ij}.
\end{split} \end{align}
Finally, when $\hat{H} = \overline{p} \hat{x}-\overline{x} \hat{p}$, the first-order moments of $e^{i \hat{H} t} \rho e^{-i \hat{H} t}$ are $\overline{r} = (\overline{x}t, \overline{p}t)^T$. This gives the correspondences:
\begin{align}
i\left[\hat{x}_i, \rho \right] \iff e_{2i},\; i\left[\hat{p}_i, \rho \right] \iff - e_{2i-1},
\end{align}
where $e_i$ is the $2n$ column vector with 1 at the $i$-th entry and all other entries 0.

The above correspondences show that the monotone metric restricted to Gaussian states is given as
\begin{widetext}
\begin{align} \begin{split} \label{GaussianMetric}
g_\rho ((d\overline{r}, dV), (d\overline{r}, dV)) &= \frac{1}{4} \sum_{i=1}^n \left[ \frac{(\text{Tr }dV\, N_i)^2}{f(1)(\nu_i^2-1)} + \frac{(\text{Tr }dV\,S_i)^2 + (\text{Tr }dV\,T_i)^2}{(\nu_i-1)^2}f\left( \frac{(\nu_i+1)^2}{(\nu_i-1)^2} \right)^{-1} \right]\\
&+ \frac{1}{8}\sum_{1\leq i<j\leq n} \left[  \frac{(\text{Tr }dV\, S_{ij})^2 + (\text{Tr }dV\, T_{ij})^2}{(\nu_i-1)(\nu_j-1)}  f\left(\frac{\nu_i+1}{\nu_i-1} \frac{\nu_j+1}{\nu_j-1} \right)^{-1} \right.\\
& \hspace{20mm}+ \left. \frac{(\text{Tr }dV\, A_{ij})^2 + (\text{Tr }dV\, B_{ij})^2}{(\nu_i+1)(\nu_j-1)}  f\left(\frac{\nu_i-1}{\nu_i+1} \frac{\nu_j+1}{\nu_j-1} \right)^{-1} \right]\\
&+ \sum_{i=1}^{n}  \frac{2}{\nu_i-1} f \left( \frac{\nu_i+1}{\nu_i-1} \right)^{-1} \left( d\overline{r}_i^2 + d\overline{r}_{i+n}^2 \right).
\end{split} \end{align}
\end{widetext}
We have used that the matrices of Eq. (\ref{mat_basis}) are orthogonal with respect to the inner product $\braket{X,Y} = \text{Tr}(X^\dagger Y)$.
Each summand of Eq. (\ref{GaussianMetric}) has a natural interpretation.
The first corresponds to change in mean photon number of each mode, the second corresponds to single-mode squeezing of each mode, the third corresponds to two-mode squeezing of each pair of modes, the fourth corresponds to beam splitting of each pair of modes, and the last corresponds to displacement of each mode.

\section{Decay Constants of GQI}\label{Sec:DC}
The results of Sec. \ref{Sec:PM} are directly applicable when one of the output states ($\rho_0$ or $\rho_1$) is a thermal state.
At target absence, the signal mode is replaced with a thermal state, so only the idler mode needs to be in a thermal state.
To directly apply the results of Sec. \ref{Sec:PM}, we first assume that the idler mode becomes a thermal state when tracing out the signal mode of the input state.
This surmounts to assuming that the input state covariance matrix satisfies $a_{33} = a_{44}$, $a_{34}=0$.

\begin{widetext}
\begin{theorem} \label{Thm:Gen}
Suppose that the input state of GQI has first-order moments $\overline{r} = (x_s, p_s, x_i, p_i)^T$ and covariance matrix $V=(a_{ij})$, satisfying $a_{33}=a_{44}=1+2N_I$ and $a_{34}=0$.
Then, the decay constants in the $\kappa \ll 1$ regime are
\begin{align}
\gamma_{\textup{col}}&=\left[ \frac{(a_{14}+a_{23})^2 + (a_{13}-a_{24})^2}{\left( \sqrt{N_B N_I} + \sqrt{(1+N_B)(1+N_I)} \right)^2}+\frac{(a_{14}-a_{23})^2 + (a_{13}+a_{24})^2}{\left(\sqrt{N_B(1+N_I)}+\sqrt{(1+N_B)N_I}\right)^2}  + \frac{8(x_s^2+p_s^2 )}{\left(\sqrt{N_B}+\sqrt{1+N_B} \right)^2}\right] \frac{\kappa}{16}, \label{gcol:gen}\\
\gamma_{\textup{loc}} &= \left[ \frac{(a_{14}+a_{23})^2 + (a_{13}-a_{24})^2}{1+N_B+N_I+2N_BN_I}+\frac{(a_{14}-a_{23})^2 + (a_{13}+a_{24})^2}{N_B+N_I+2N_BN_I}  + \frac{8(x_s^2+p_s^2)}{1+2N_B} \right] \frac{\kappa}{32}. \label{gloc:gen}
\end{align}
\end{theorem}
\end{widetext}
The assumptions $a_{33}=a_{44}=1+2N_I$ and $a_{34}=0$ are not that restrictive, since in QI, we are able to perform any local unitary on the idler mode without affecting the decay constants.
Therefore, Thm. \ref{Thm:Gen} is actually applicable to any two-mode Gaussian state.
The symplectic diagonalization of a single-mode state can be written analytically, so we can also write an analytic formula of the decay constants  $\gamma_{\text{col}}$ and $\gamma_{\text{loc}}$ for arbitrary two-mode Gaussian states.
However, the formula of symplectic diagonalization is lengthy, which leads to an incomprehensible equation for the decay constants.
In the limit of large thermal background, $N_B \gg 1$, the formula for arbitrary two-mode Gaussian states becomes tractable as follows.
\begin{theorem}
Suppose that the input state of GQI has first-order moments $\overline{r} = (x_s, p_s, x_i, p_i)^T$ and covariance matrix $V= \begin{pmatrix} V_s & C \\ C^T & V_i \end{pmatrix}$ where $V_s$, $V_i$, $C$ are $2\times 2$ matrices.
Then, in the $\kappa \ll 1 \ll N_B$ regime, we have
\begin{align}
\gamma_{\textup{col}} &= \frac{\kappa}{4N_B} \left(\frac{x_s^2+p_s^2}{2} + \frac{\nu_2 \textup{Tr }CV_i^{-1} C^T}{(\sqrt{\nu_2+1}+\sqrt{\nu_2-1})^2} \right),\label{gcol:genNB}\\
\gamma_{\textup{loc}} &= \frac{\kappa}{4N_B} \left(\frac{x_s^2+p_s^2}{2} + \frac{\textup{Tr }CV_i^{-1} C^T}{4} \right).
\end{align}
Here, $\nu_2 = \sqrt{\textup{det } V_i}$.
\end{theorem}
The importance of this form is that we do not need any information of the symplectic diagonalization of the covariance matrix.
Another important feature is the simple $\kappa/N_B$ dependence.
From now on, the graphs we consider show the decay constants normalized with respect to some decay constant with same $\kappa$ and $N_B$, so the graphs are independent of $\kappa$ and $N_B$ provided $\kappa \ll 1 \ll N_B$.
The details of the $N_B\gg1$ limit is discussed further in Appendix \ref{App:Lim}.
We performed a numerical verification that for $\kappa = 0.01$, $N_B=625$, Eq. (\ref{gcol:gen}, \ref{gcol:genNB}) provide accurate approximations to the quantum Chernoff bound Eq. (\ref{QCB:def}).

\subsection{Comparison with Previous Works}
Thm. \ref{Thm:Gen} reproduces the results which were already known.
If the input state is a coherent state with signal mean photon number $N_S$, we have
\begin{equation} \label{eqn:CI}
\gamma_{\text{col}}^{\text{CI}}= \frac{N_S \kappa}{\left(\sqrt{N_B}+\sqrt{1+N_B} \right)^2},\;\; \gamma_{\text{loc}}^{\text{CI}} = \frac{N_S \kappa}{2+4N_B}.
\end{equation}
If the input state is a TMSV state with signal mean photon number $N_S$ we have
\begin{align}
\gamma_{\text{col}}^{\text{TMSV}} &=\frac{N_S (1+N_S)\kappa}{\left( \sqrt{N_B N_S} + \sqrt{(1+N_B)(1+N_S)} \right)^2},\\
\gamma_{\text{loc}}^{\text{TMSV}} &= \frac{N_S(1+N_S)\kappa}{2+2N_B+2N_S+4N_B N_S}.
\end{align}
In the usual limit $N_S\ll 1 \ll N_B$, we see that $\gamma_{\text{col}}^{\text{TMSV}} = 4 \gamma_{\text{col}}^{\text{CI}}$ and $\gamma_{\text{loc}}^{\text{TMSV}} = 2\gamma_{\text{loc}}^{\text{CI}}$ recovering the 6 dB advantage of collective measurements \cite{ST08} and 3 dB advantage of local measurements \cite{MS17}.

The input states with arbitrary TMSV-like correlation were recently considered in Ref. \cite{AK20}.
When using the input state with zero first-order moments and covariance matrix
\begin{equation}
V = \begin{pmatrix} a&&2c& \\ &a&&-2c \\2c &&a& \\ &-2c&&a \end{pmatrix},
\end{equation}
$a=1+2N_S$, and $|c|\leq \sqrt{N_S+N_S^2}$, it was numerically shown that $\gamma_{\text{col}} = \kappa c^2 / N_B $ in the $N_S \ll 1 \ll N_B$ regime.
Eq. (\ref{gcol:gen}) yields
\begin{equation}
\gamma_{\text{col}} =\kappa \left( \frac{ c}{\sqrt{N_B N_S}+\sqrt{(1+N_B)(1+N_S)}} \right)^2
\end{equation}
which reproduces the numerical result in the $N_S \ll 1 \ll N_B$ regime.

\begin{figure}[t]
\includegraphics[width=0.49\textwidth]{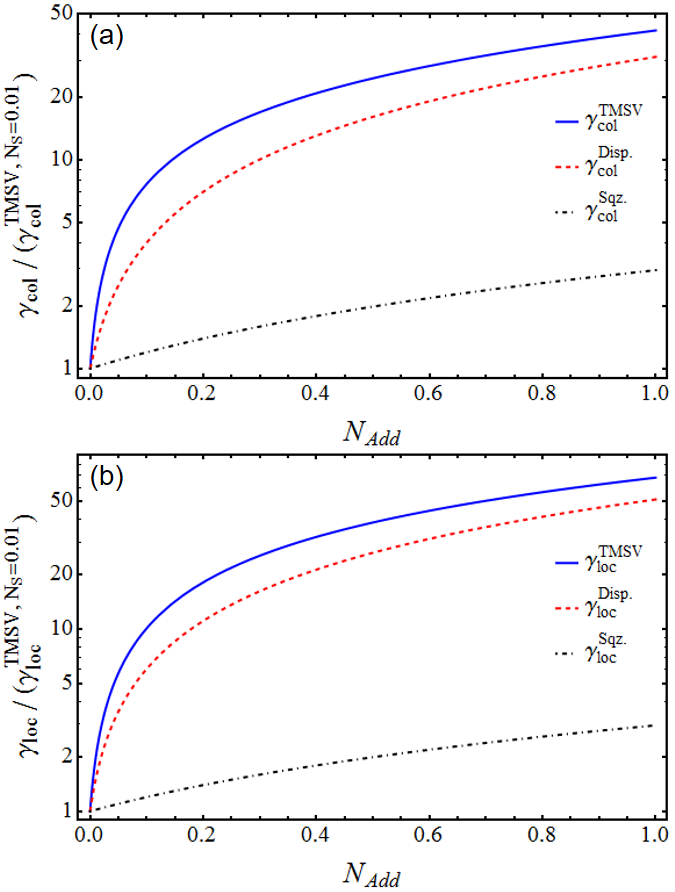}
\caption{Graph of decay constants of (a) collective measurements and (b) local measurements as a function of added signal mean photon number when we apply additional Gaussian operations on the TMSV state with $N_S=0.01$.
Decay constants are normalized with respect to the decay constant of TMSV state with $N_S=0.01$.
Producing TMSV states with higher mean photon number (blue, solid) is optimal, and performing displacement on signal mode (red, dashed) outperforms performing single-mode squeezing on signal mode (black, dot-dashed).
Thermal background has mean photon number $N_B=625$.}
\label{fig:sigop}
\end{figure}

\begin{figure*}[t]
\includegraphics[width=0.99\textwidth]{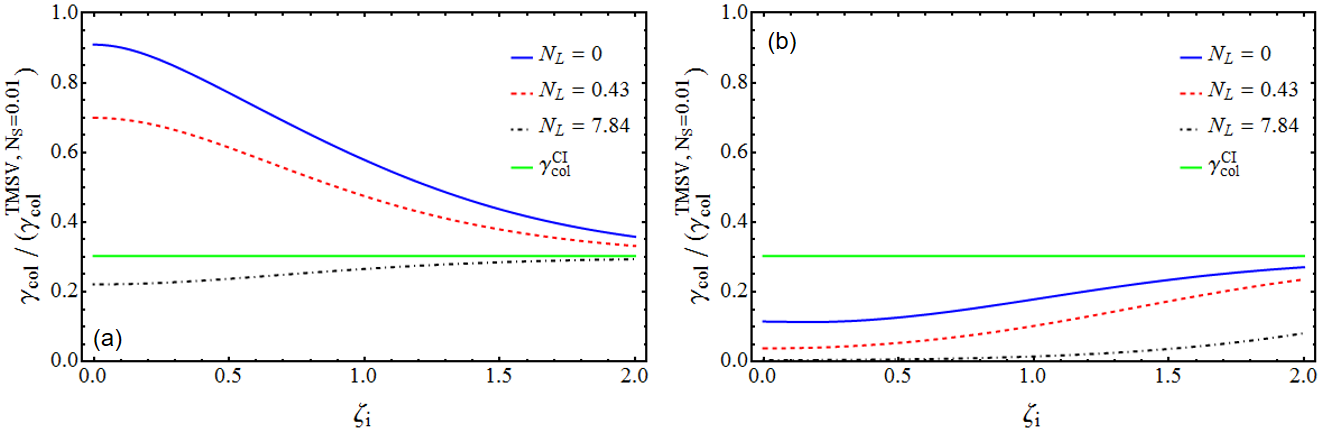}
\caption{Graph of collective decay constant $\gamma_{\text{col}}$ under (a) idler memory transmittivity $\eta=0.9$, (b) $\eta=0.1$ as a function of idler squeezing on TMSV state with $N_S=0.01$ before loss.
Values are normalized with respect $\gamma_{\text{col}}$ of TMSV state with $N_S=0.01$.
The idler memory noise is $N_L=0$ (blue, solid), $N_L=0.43$ (red, dashed), and $N_L=7.84$ (black, dot-dashed) which corresponds to the thermal noise of frequency 10 GHz at temperature 40 mK, 400 mK, and 4 K respectively.
The green solid line is the decay constant of coherent state with equal signal mean photon number.
Thermal background has mean photon number $N_B=625$.
Local decay constants $\gamma_{\text{loc}}$ show the same qualitative behavior.}
\label{fig:idlsqz}
\end{figure*}

\subsection{Optimal and Sub-optimal Input States}
The optimality of TMSV states in QI was investigated in various contexts \cite{GP18, RN20, MB21}.
We find optimal states which maximize $\gamma_{\text{col(loc)}}$ under the constraint of fixed signal mean photon number as in Ref. \cite{MB21}.
Among pure Gaussian states with fixed signal mean photon number, we show in Appendix \ref{App:Opt} that TMSV states maximize both decay constants.
Therefore, the best strategy to improve performance of GQI is to increase the two-mode squeezing parameter of the entangled source.
Our proof also provides a reason to why the quantum advantage is prominent in the low signal mean photon number regime.

However, when it is not possible to increase the two-mode squeezing parameter, we may instead consider additional single-mode Gaussian operations on the signal mode of a TMSV state to increase the decay constants.
In Fig. \ref{fig:sigop}, we compare the effect of displacement and single-mode squeezing on the signal mode on decay constants.
The phase of single-mode squeezing is aligned with the two-mode squeezing phase to maximize decay constants.
Both operations increase signal mean photon number, hence we observe an increase in decay constants.
Since we have shown that TMSV states are optimal among states with same signal mean photon number, either applying displacement or single-mode squeezing on the signal mode cannot outperform the TMSV state with same signal mean photon number.
We see that displacement is beneficial than single-mode squeezing on signal mode for both decay constants.
This is true regardless of the initial signal mean photon number of the TMSV source or thermal background.
Experimental implementation of displacement \cite{MP96, KF16} is much more easier than that of single-mode squeezing, as the latter requires nonlinear interactions with feed-forward  or heralding methods \cite{BY89, RF05, JZ20}.
Hence, signal-mode displaced TMSV states will be the realistically sub-optimal state to prepare in experiments.
Note that with an ideal idler memory, single-mode Gaussian operations on the idler mode will have no effect on decay constants, as we can always apply the inverse operation on the idler and restore the original state.

\section{GQI with Realistic Idler Memory}\label{Sec:Loss}
In this section, we study the performance of GQI under the presence of a realistic idler memory channel with loss and noise.
This memory channel is modeled as a Gaussian thermal loss channel which mixes the input and a thermal state through a beam splitter.
In terms of Fig. \ref{Fig:QI} (c), we take $\Phi_{sig}$ to be the single-mode identity channel and $\Phi_{id}$ to be the single-mode Gaussian channel which acts as $\overline{r}\mapsto\sqrt{\eta} \overline{r}$ on first-order moments and $V\mapsto\eta V + (1-\eta)N_L$ on covariance matrix.
$\eta$ is the transmittivity of the beam splitter and $N_L$ is the mean photon number of the thermal state.
We first investigate the effect of single-mode Gaussian unitaries on the idler mode before the memory in order to compensate loss and noise, so that $\Phi_{id}$ consists of a unitary and the aforementioned memory channel.
It turns out that actually, no single-mode Gaussian unitary channel can help us to both compensate loss and beat coherent states.
This underlines the importance of preparing a memory channel close to ideal.
We then compare signal-mode displaced TMSV states against coherent states to quantify the conditions on the idler memory required to observe quantum advantage over coherent states.

\subsection{Gaussian Unitaries on Idler Mode} \label{SSec:Limit}
One would try to compensate the loss of the idler mode by performing single-mode Gaussian unitaries on the idler mode before the memory channel.
The first-order moments of the idler mode do not appear in Thm. \ref{Thm:Gen}, hence displacing the idler mode will have no effect on the decay constants.
We can understand this physically: if we displace the idler mode by $r_i$ before the memory channel and then displace by $-\sqrt{\eta}r_i$ after, this is equivalent to doing nothing on the source, hence the decay constants are independent of idler displacement.

\begin{figure*}[t]
\includegraphics[width=0.99\textwidth]{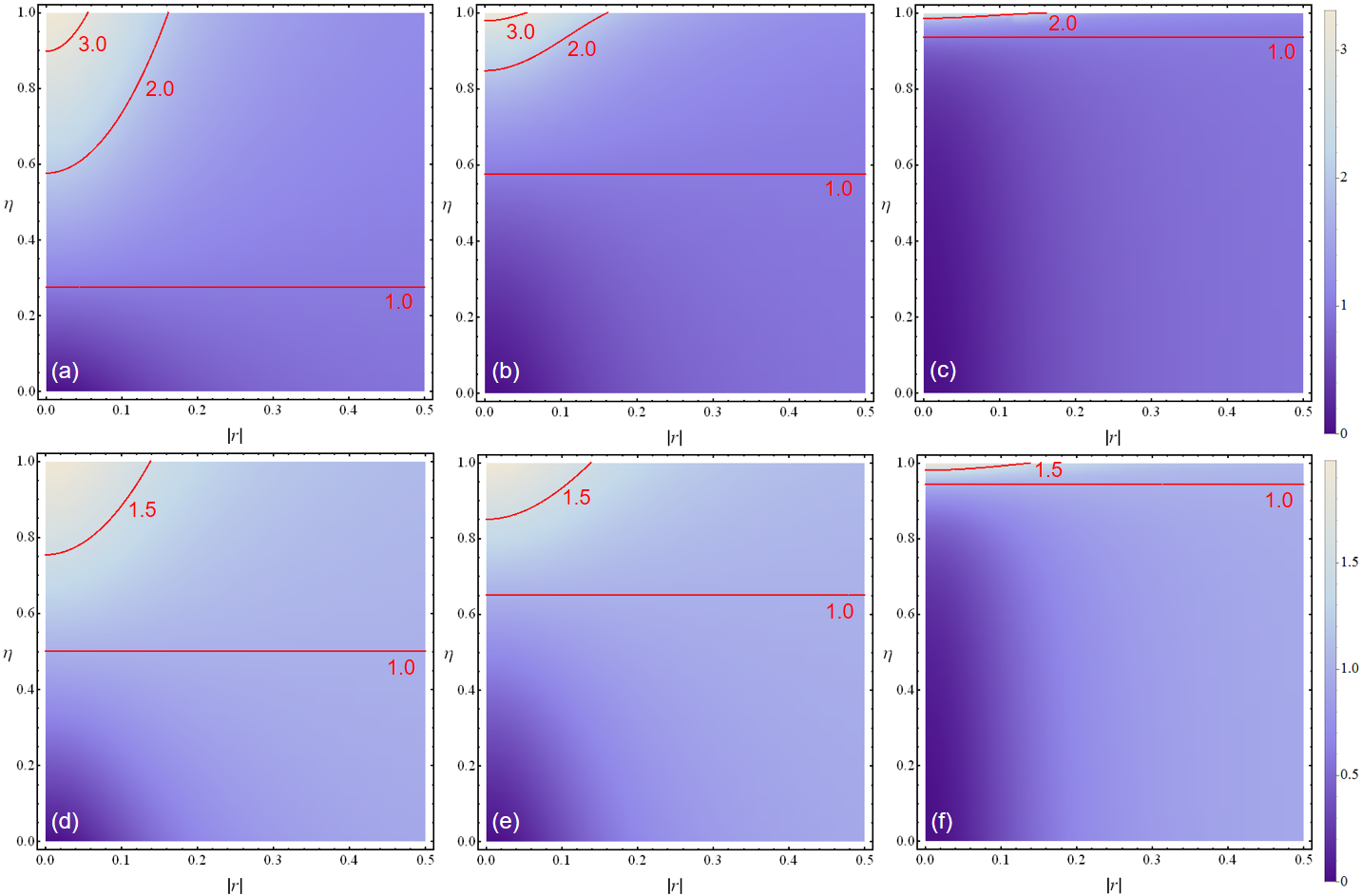}
\caption{The quantum advantage of (a, b, c) collective measurements $QA_{\text{col}}$ or (d, e, f) local measurements $QA_{\text{loc}}$ of displaced TMSV states as a function of signal-mode displacement parameter $|r|$ and idler memory transmittivity $\eta$ at (a, d) $N_L = 0$, (b, e) $N_L=0.43$, and (c, f) $N_L=7.84$.
The TMSV state has signal mean photon number $N_S=0.01$ and thermal background is $N_B=625$.
The contour of $QA_{\text{col(loc)}}=1$ is independent of signal-mode displacement parameter $|r|$.}
\label{fig:QA}
\end{figure*}

Fig. \ref{fig:idlsqz} shows how $\gamma_{\text{col}}$ changes as a function of idler squeezing $\zeta_i$ applied to a TMSV state under various combinations of idler transmittance $\eta$ and noise $N_L$.
The phase of idler squeezing is aligned with the two-mode squeezing phase to maximize decay constants.
Specifically, when we use a TMSV state with covariance matrix given as Eq. (\ref{covmat:TMSV}), the corresponding symplectic matrix of idler squeezing is $\text{diag}(1,1,e^{\zeta_i}, e^{-\zeta_i})$.
Whether performing squeezing on the idler mode is beneficial or not depends on the values of $\eta$ and $N_L$.
Idler squeezing is beneficial only when the idler channel has high loss or noise.
The decay constants actually converge as we send idler squeezing to infinity, values given as
\begin{align}
\lim_{\zeta_i \rightarrow \infty} \gamma_{\text{col}} &= \frac{N_S \kappa}{(\sqrt{N_B}+\sqrt{1+N_B})^2}\frac{1+N_S}{1+2N_S},\\
\lim_{\zeta_i \rightarrow \infty} \gamma_{\text{loc}} &= \frac{N_S \kappa}{2+4N_B} \frac{1+N_S}{1+2N_S}.
\end{align}
Direct comparison with the decay constants of coherent states, Eq. (\ref{eqn:CI}), shows that the limits are always smaller than the decay constants of coherent states.
The main reason that we cannot beat coherent states in this scheme is that the state after idler loss and before signal-target interaction is separable when idler squeezing is high enough.
Idler squeezing amplifies one of the correlations while de-amplifying the other and we cannot recover the de-amplified correlation after idler loss, making the state after idler loss separable \cite{RS00}.
We will discuss more about separable states in Sec. \ref{Sec:Corr}.

The limit of decay constants at infinite idler squeezing actually exists for any Gaussian state. We have the following theorem.
\begin{theorem} \label{Thm:Loss}
Suppose that the input state of GQI has first-order moments $\overline{r} = (x_s, p_s, x_i, p_i)^T$ and covariance matrix $V=(a_{ij})$ satisfying $a_{34}=0$.
The idler mode undergoes squeezing and a memory channel.
Then, the decay constants at infinite idler squeezing are
\begin{align} 
\lim_{\zeta_i \rightarrow \infty} \gamma_{\textup{col}}&=\left[ \frac{(a_{13}^2+a_{23}^2)/a_{33}+2(x_s^2+p_s^2 )}{4\left(\sqrt{N_B}+\sqrt{1+N_B}\right)^2} \right] \kappa, \\
\lim_{\zeta_i \rightarrow \infty} \gamma_{\textup{loc}} &= \left[ \frac{(a_{13}^2+a_{23}^2)/a_{33}+2(x_s^2+p_s^2 )}{8+16N_B}  \right] \kappa.
\end{align}
The limits are independent of characteristics of the idler memory, $\eta$ and $N_L$, and always less than the decay constants of coherent states.
\end{theorem}
That coherent states beat the limit follows directly from the Cauchy--Schwarz inequality $|\braket{\hat{a}_s \hat{x}_i}|^2 \leq \braket{\hat{a}^\dagger_s \hat{a}_s} \braket{\hat{x}_i^2}$.
This implies that given any Gaussian state that beats the coherent state with same signal mean photon number, squeezing the idler mode to compensate idler loss will actually lead to worse performance.
These results stress the importance of a lossless and noiseless quantum memory to see quantum advantage in QI.

\subsection{Idler Memories with Quantum Advantage}
The importance of preparing an efficient quantum memory was pointed out in the previous section.
We define the quantum advantage of a state as the ratio of the decay constant of that state to the decay constant of the coherent state with same signal mean photon number, $QA_{\text{col(loc)}} = \gamma_{\text{col(loc)}} / \gamma_{\text{col(loc)}}^{\text{CI}}$.
Fig. \ref{fig:QA} shows the quantum advantage of collective or local measurements when using displaced TMSV states as varying the idler memory characteristics.
The quantum advantage decreases as we increase loss or noise in the idler memory as one would expect.
As we increase the signal-mode displacement parameter $|r|=(x_s^2+p_s^2)^{1/2}$, the coherent contribution to decay constants surpasses the contribution from TMSV-like correlations.
This means that the quantum advantage will become unity as we increase the displacement parameter.
In other words, there is a limit on displacement also to see a quantum advantage strictly larger than unity as shown in the contours of Fig. \ref{fig:QA}.

The contour of $QA_{\text{col(loc)}}=1$ is independent of displacement, which mainly stems from the fact that the signal mean photon number of displaced TMSV states is simply the sum $N_S+|r|^2/2$.
This contour provides a criterion for idler memory loss in order to observe quantum advantage.
We need to prepare a quantum memory with transmittivity larger than $\eta_{\text{col(loc)}}^{QA=1}$, the transmittivity which achieves $QA_{\text{col(loc)}}=1$.
$\eta_{\text{col(loc)}}^{QA=1}$ is a function of $N_S, N_B, N_L$.
For local measurements, we have
\begin{equation}
\eta_{\text{loc}}^{QA=1} = \frac{1+N_B+N_L+2N_B N_L}{(1+2N_B)(1+N_L)}>\frac{1}{2}.
\end{equation}
The analytic formula for $\eta_{\text{col}}^{QA=1}$ is much more complicated.
However, by comparing $\gamma_{\text{col}}$ of a TMSV state with idler transmittivity $1/4$ and a coherent state, we have $\eta_{\text{col}}^{QA=1} > 1/4$.
Hence to have quantum advantage with collective (resp. local) measurements, we need to prepare a quantum memory with transmittivity higher than $1/4$ (resp. $1/2$).
The decay constants of GQI with a noiseless but lossy idler memory using TMSV states in the $N_S\ll 1 \ll N_B$ limit are $\gamma_{\text{col}} = \eta\kappa N_S/N_B$, $\gamma_{\text{loc}} = \eta \kappa N_S/2N_B$, so the 6 dB (resp. 3 dB) advantage of collective (resp. local) measurements are converted to a minimum transmittivity of $1/4$ (resp. $1/2$).

\begin{figure}[t]
\includegraphics[width=0.49\textwidth]{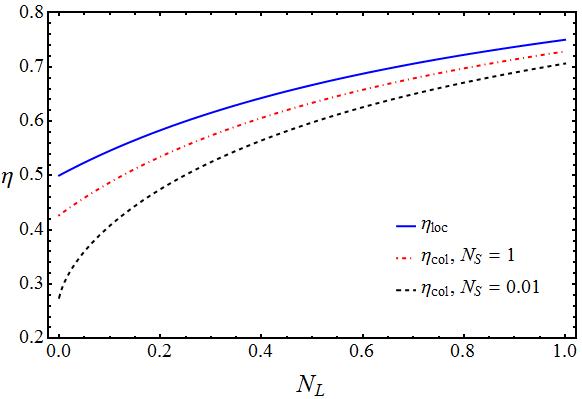}
\caption{The minimal transmittivity to observe quantum advantage $\eta_{\text{col(loc)}}^{QA=1}$ in the $N_B\rightarrow \infty$ limit as a function of idler memory noise $N_L$ and TMSV signal mean photon number $N_S$.
The requirement on transmittivity when using local measurements is independent of $N_S$.
The requirement when using collective measurements decreases as $N_S$ decreases, so in the low signal power regime, we have more flexibility in allowed idler memory transmittivity to have quantum advantage.}
\label{fig:TC}
\end{figure}

Instead of giving the full form of $\eta_{\text{col}}^{QA=1}$, we provide formulae at the $N_B \rightarrow \infty$ limit:
\begin{align}
\lim_{N_B \rightarrow\infty}& \eta_{\text{loc}}^{QA=1} =\frac{1}{2} \frac{1+2 N_L}{1+N_L},\\
\begin{split}
\lim_{N_B \rightarrow\infty} &\eta_{\text{col}}^{QA=1} =\frac{1}{4} \frac{1+2 N_L}{1+N_L}\\
&\times\left(1+\sqrt{1-\frac{1+N_L}{(1+2N_L)^2(1+N_S)}} \right).
\end{split}
\end{align}
Fig. \ref{fig:TC} shows the dependency of $\eta_{\text{col(loc)}}^{QA=1}$ as a function of $N_L$ and $N_S$.
Both values increase to unity as $N_L$ increases, implying that noise in the idler memory puts more stringent conditions on the required idler transmittivity.
The minimum transmittivity for local measurements is independent of $N_S$, while for collective measurements, it is not.
For example, if $N_L=0.43$, a quantum memory with transmittivity larger than 65\% is needed regardless of $N_S$ when using local measurements.
When we are able to perform optimal collective measurements and use a TMSV source with signal mean photon number $N_S=1$, we can lower the requirement to a quantum memory with transmittivity larger than 61.5\%.
If we reduce the signal mean photon number to $N_S=0.01$, we can further lower the requirement to a quantum memory with transmittivity larger than 57.5\%.
The decrease of the transmittivity requirement is more dramatic with smaller $N_L$.

\begin{figure*}[t]
\includegraphics[width=0.99\textwidth]{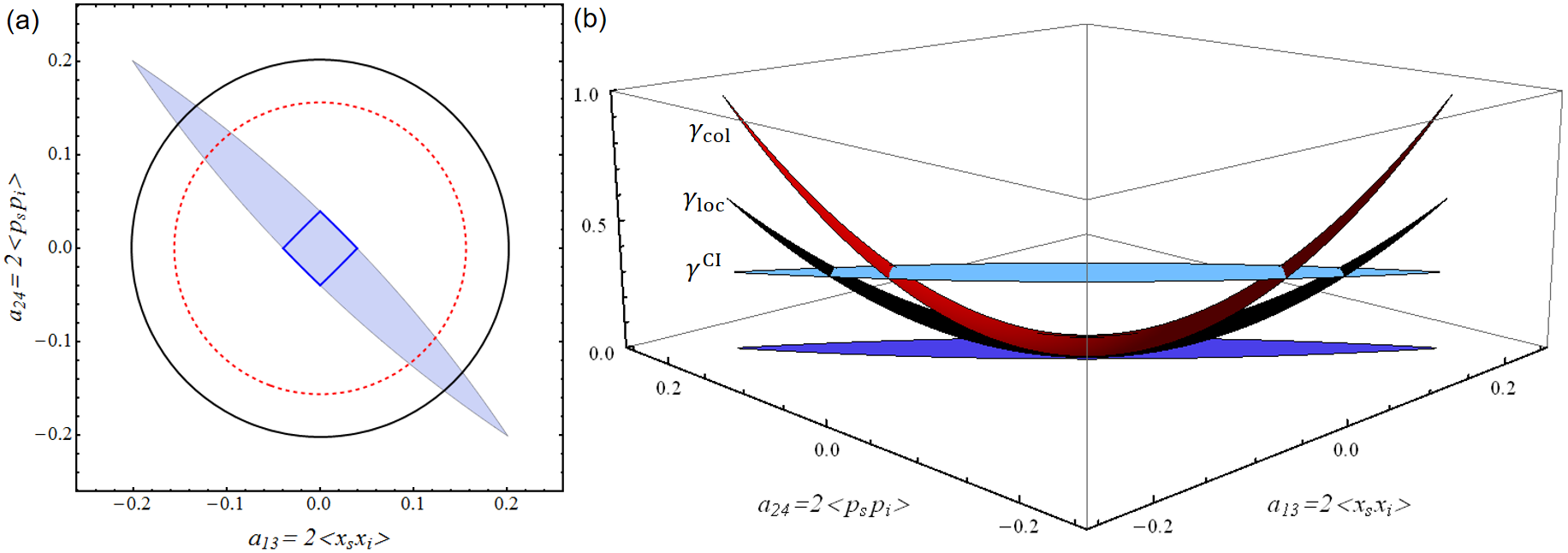}
\caption{(a) The shaded region shows the quantum mechanically allowed correlations of states with covariance matrix given as Eq. (\ref{Vcorr}) under the constraint $N_S=N_I=0.01$.
The outmost states along the anti-diagonal correspond to TMSV states, the origin corresponds to thermal state.
The inner diamond (blue, solid) encloses separable states.
The middle (resp. outer) circle (red, dashed (resp. black, solid)) is the contour of correlations whose decay constant $\gamma_{\text{col}}$ (resp. $\gamma_{\text{loc}}$) equals that of a coherent state with same signal mean photon number.
Thermal background has mean photon number $N_B=625$.
States outside the black contour can outperform coherent states through local measurements.
States between the black and red contour can outperform coherent states only by collective measurements.
States inside the red contour cannot outperform coherent states even when using the optimal measurement.
(b) 3D plot of the decay constants $\gamma_{\text{col}}$ and $\gamma_{\text{loc}}$ as a function of correlations $a_{13}$ and $a_{24}$ restricted to the region of physical correlations under the same conditions of (a).
$\gamma^{\text{CI}}$ is the decay constant when using a coherent state of mean photon number $N_S=0.01$.
The intersection of the graph of $\gamma^{\text{CI}}$ and the graph of $\gamma_{\text{col}}$ (resp. $\gamma_{\text{loc}}$) is the red (resp. black) contour of (a).
}
\label{fig:Corr}
\end{figure*}

\section{Correlations Beating Coherent States}\label{Sec:Corr}
We are interested in values of covariance matrix elements $a_{13}, a_{14}, a_{23}, a_{24}$ describing correlations between the signal and idler which can beat coherent states under the constraint of same signal mean photon number.
It is evident from the decay constants Eq. (\ref{gcol:gen}, \ref{gloc:gen}) that if we can increase all correlation covariance matrix elements by the same ratio, then the decay constant will also increase.
However, when the signal and idler mean photon numbers are fixed, the uncertainty principle places bounds on the physically possible values of correlation covariance matrix elements.
For simplicity, we will consider states with covariance matrix of form
\begin{equation} \label{Vcorr}
V=\begin{pmatrix} 1+2N_S && a_{13}& \\ & 1+2N_S&&a_{24} \\ a_{13}&&1+2N_I & \\ &a_{24} &&1+2N_I \end{pmatrix} 
\end{equation}
with zero first-order moments.
Fig. \ref{fig:Corr} (a) shows the region of quantum mechanically allowed correlations $a_{13}$ and $a_{24}$ in this case.
The origin corresponds to the thermal state, the points furthest from the origin on the anti-diagonal correspond to TMSV states, and the points furthest from the origin on the diagonal correspond to a beam split thermal state.
The circle contours of Fig. \ref{fig:Corr} (a) show the correlations which satisfy $\gamma_{\text{col}} = \gamma_{\text{col}}^{\text{CI}}$ (red, dashed) or $\gamma_{\text{loc}} = \gamma_{\text{loc}}^{\text{CI}}$ (black, solid) including correlations which violate the uncertainty principle.
The decay constants increase as we move further from the origin, hence the shaded region outside the circle contours represent physical states which show quantum advantage.
We see that it is not necessary to ideally prepare TMSV states in order to observe quantum advantage.

The inner diamond of Fig. \ref{fig:Corr} (a) shows the boundary of separable states.
As separable states are inside the contours, we see that separable states cannot show advantage over coherent states.
We can argue that in general, separable states will have a decay constant less than the optimal single-mode state as follows.
Using product states $\rho_s \otimes \rho_i$ as the input for QI is essentially equivalent to using a single-mode input $\rho_s$ to QI, hence cannot beat the optimal single-mode probe state.
The decay constant $\gamma_{\text{col}}$ is jointly convex \cite{KA07}, implying that separable states cannot also beat the optimal single-mode probe state.
In terms of $\gamma_{\text{col}}$, coherent states are the optimal single-mode probe state \cite{MB21}.
(The analysis of \cite{MB21} is incomplete as they only show that coherent states are critical points.
A complete analysis should find all critical points and maximize among the critical points.
In Appendix \ref{App:SMQI}, we show that the only critical points are displaced number states $\hat{D}_{\overline{r}} \ket{n}$ and coherent states indeed maximize$\gamma_{\text{col}}$.)
Therefore, separable states cannot outperform coherent states, explaining the phenomenon occurring at infinite idler squeezing discussed in Sec. \ref{SSec:Limit}.

\section{Conclusions} \label{Sec:Con}
We evaluated monotone metrics on Gaussian states and showed that the metrics are naturally diagonalized in terms of symplectic eigenvalues, single- and two-mode squeezing, beam splitting, and displacement parameters.
This is relevant to QI, since the decay constants of QI become special cases of monotone metrics under the assumption of low target reflectivity.
We applied our results to obtain analytic formulae for decay constants of QI in terms of the first-order moments and the covariance matrix of the input state, provided some constraint on the idler mode.
For general input states, we only need the symplectic diagonalization of the idler mode of the input state, while previous methods required the symplectic diagonalization of the two output two-mode states.
Beyond applications to QI, any quantum divergence which is monotone under CPTP maps becomes a monotone metric in its infinitesimal form, so we can apply our results to the computation of these quantities for Gaussian states.

Based on these results, we first showed that TMSV states are optimal in GQI among pure Gaussian states with same signal power.
Then we considered the effect of displacement and single-mode squeezing on the signal mode of a TMSV state.
As both operations increase signal mean photon number, the decay constants increase as expected.
We showed that signal-mode displaced TMSV states are a sub-optimal probe, which is advantageous as displacement is experimentally more feasible than single-mode squeezing.
The effect of idler loss and methods to compensate it were also considered.
Single-mode squeezing on the idler mode can increase the QI performance when the idler loss is large enough, but unfortunately in this regime, the loss is so large that we cannot beat coherent states.
This implies the importance of preparing an efficient idler memory.
We also provided the criteria on idler loss to see quantum advantage over coherent states in terms of other system parameters.
Finally, we showed the region of correlations a two-mode state must have to beat coherent states.

\begin{acknowledgments}
This work was supported by a grant to Defense-Specialized Project funded by Defense Acquisition Program Administration and Agency for Defense Development.
\end{acknowledgments}

\appendix
\section{Bargmann Representation}\label{App:BR}
We give a brief introduction to the Bargmann representation, which provides a compact method to deal with bosonic systems.
For a single-mode bosonic system, the basic correspondence is 
\begin{equation}
\ket{n} \iff \frac{1}{\sqrt{n!}} z^n,\; \bra{n} \iff \frac{1}{\sqrt{n!}} w^{*n}.
\end{equation}
Then an operator $\hat{O} = \sum_{n,m} O_{nm} \ket{n} \bra{m}$ corresponds to
\begin{equation}
K_O (z,w^*) = \sum_{n,m}\frac{O_{nm}}{ \sqrt{n!m!}} z^n w^{*m}.
\end{equation}
The simple integral
\begin{equation}
\frac{1}{\sqrt{n!m!}} \int_{\mathbb{C}} \frac{d^2 z}{\pi}\; z^n z^{*m} e^{-|z|^2} = \delta_{n,m}
\end{equation}
where $d^2 z = d(\text{Re}z)\; d(\text{Im}z)$ implies that the trace and composition is given as
\begin{align}
\text{Tr } \hat{O} &= \int_{\mathbb{C}} \frac{d^2z}{\pi}\; K_O(z,z^*) e^{-|z|^2},\\
K_{O_1 O_2} (z, w^*) &= \int_{\mathbb{C}} \frac{d^2\alpha}{\pi} \; K_{O_1} (z,\alpha^*) K_{O_2} (\alpha, w^*) e^{-|\alpha|^2}.
\end{align}
Finally, to reconstruct the Fock basis element $O_{nm}$, one needs to know the derivatives of $K_O$ at 0:
\begin{equation}
O_{nm} = \frac{1}{\sqrt{n!m!}} \left. \left( \frac{\partial}{\partial z} \right)^n \left( \frac{\partial}{\partial w^*} \right)^m K_O \right|_{z=w=0}.
\end{equation}
The extension to multi-mode cases is trivial.
More details on the Bargmann representation can be found in Ref. \cite{BH00} or chapter 7 of Ref. \cite{JK68}.

\section{Decay Constants under Large Background} \label{App:Lim}
The thermal background at microwave frequencies is too large, while in optical frequencies it is negligible.
For example, the mean photon number of a thermal state at 300 K, 10 GHz is about $N_B \simeq 625 \gg 1$.
Here, we focus on the $N_B\gg 1$ regime and show that the formulae for decay constants for arbitrary two-mode Gaussian states are simplified in the high background regime.

Let the covariance matrix of the input state be
\begin{equation}
V= \begin{pmatrix} V_s & C \\ C^T & V_i \end{pmatrix}
\end{equation}
so $V_{s(i)}$ is the covariance matrix of the signal (idler) when the other mode is traced out.
Furthermore, let $V_i = \nu_2 SS^T$ be the symplectic diagonalization of $V_i$ for some $S\in Sp(1)$.
Then, the monotone metric between the two output states of QI is
\begin{align}
&g_{\rho_0}(d\rho, d\rho) \nonumber \\
&= \frac{1}{8} \left[  \frac{(\text{Tr }dV\, S_{12})^2 + (\text{Tr }dV\, T_{12})^2}{(\nu_1-1)(\nu_2-1)}  f\left(\frac{\nu_1+1}{\nu_1-1} \frac{\nu_2+1}{\nu_2-1} \right)^{-1} \right. \nonumber \\
& + \left. \frac{(\text{Tr }dV\, A_{12})^2 + (\text{Tr }dV\, B_{12})^2}{(\nu_1+1)(\nu_2-1)}  f\left(\frac{\nu_1-1}{\nu_1+1} \frac{\nu_2+1}{\nu_2-1} \right)^{-1} \right]\nonumber \\
& + \frac{2}{\nu_1-1} f \left( \frac{\nu_1+1}{\nu_1-1} \right)^{-1}r^2, \\
&dV = \begin{pmatrix} & CS^{-T} \\ S^{-1}C^T& \end{pmatrix},
 \end{align}
where $d\rho = (d\overline{r}, dV)$ as in Eq. (\ref{dQI}), $\nu_1 = 1+2N_B$, and $r^2 =x_s^2 + p_s^2$.
In the $N_B\gg1$ regime, this becomes
\begin{align} \begin{split}
g_{\rho_0}&(d\rho, d\rho)=\frac{1}{16N_B(\nu_2-1)}f \left(\frac{\nu_2+1}{\nu_2-1} \right)^{-1}\\
& \times \left[ \text{Tr}(dV\, S_{12})^2 + \text{Tr}(dV\, T_{12})^2 \right. \\
& \hspace{20mm} \left. +\text{Tr}(dV\, A_{12})^2 + \text{Tr}(dV\, B_{12})^2 \right]\\
&+ \frac{1}{N_B f(1)}r^2.
\end{split} \end{align}
The sum in the square brackets can be simplified as
\begin{align} \begin{split}
 &\left[ (\text{Tr }dV\, S_{12})^2 + (\text{Tr }dV\, T_{12})^2 +(\text{Tr }dV\, A_{12})^2 \right. \\
& \hspace{2mm} \left.  + (\text{Tr }dV\, B_{12})^2 \right]= 8 \nu_2 \text{Tr }CV_i^{-1} C^T.
\end{split} \end{align}
Therefore, we have
\begin{align} \begin{split}
\kappa g_{\rho_0}(d\rho, d\rho) &= \frac{\kappa}{N_B} \left[ \frac{r^2}{f(1)}+ \frac{\nu_2}{2(\nu_2-1)} \right.\\
&\times \left. f \left(\frac{\nu_2+1}{\nu_2-1} \right)^{-1} \text{Tr }CV_i^{-1} C^T \right].
\end{split} \end{align}
The simple $\kappa/N_B$ dependence means that it is unnecessary to specify the values of $\kappa$ and $N_B$ when drawing the graphs normalized to a specific value provided $\kappa \ll 1 \ll N_B$.
The decay constants $\gamma_{\text{col(loc)}}$ are obtained by using the specific functions $f_{\text{col}}(t) = 2(\sqrt{t}+1)^2$ or $f_{\text{loc}}(t) = 4(t+1)$.
We have obtained a concise way to compute the decay constants of QI for an arbitrary two-mode Gaussian state in the $N_B\gg1$ regime.

\section{Optimal Two Mode Input of GQI} \label{App:Opt}
We will prove the following theorem.
\begin{theorem}
Among pure Gaussian states with fixed signal mean photon number, TMSV states maximize $g_{\rho_0}(d\rho, d\rho)$ where $d\rho = (d\overline{r}, dV)$ as in Eq. (\ref{dQI}) and $g$ is any monotone metric.
Hence TMSV states maximize both decay constants $\gamma_{\text{col}}$ and $\gamma_{\text{loc}}$ among pure Gaussian states with equal signal  mean photon number.
\end{theorem}
From our assumptions, we can write the covariance matrix as $V= SS^T$ where $S\in Sp(2)$. Before using Lagrange multipliers, we reduce the number of variables as much as possible.

As the idler mode is retained ideally, applying any local unitary on the idler mode does not affect the quantities of interest.
Hence we may assume that the idler mode is thermal when tracing out the signal mode, i.e., $a_{33} = a_{44}$ and $a_{34}=0$.
The two channels of target present and absent are both insensitive to local phase rotations of signal and idler modes.
By rotating the signal mode phase, we can take $a_{12}=0$. By rotating the idler mode phase, we can take $a_{14}=0$.
By the following lemma, we can take $a_{23}=0$.

\begin{lemma}
Let $V=(a_{ij})$ be the covariance matrix of a two-mode pure Gaussian state. Suppose $a_{12}=a_{14} = a_{34}=0$. Then, $a_{23}=0$.
\end{lemma}
\textit{Proof}. We may write $V=SS^T$ with $S\in Sp(2)$. Write $S$ as
\begin{equation}
S = \begin{pmatrix}
&\bra{1}& \\ &\bra{2}& \\ &\bra{3}& \\ &\bra{4}&
\end{pmatrix}
\end{equation}
so $\bra{i}$ are row vectors of $S$.
The assumptions on matrix elements become $\braket{1|2} = \braket{1|4} = \braket{3|4}=0$.
$S\in Sp(2)$ implies $\braket{1|\Omega|3}=\braket{1|\Omega|4}=\braket{2|\Omega|4}=0$.
The four vectors $\ket{2}, \ket{4}, \Omega\ket{3}, \Omega\ket{4}$ are all perpendicular to $\bra{1}$, so $\Omega\ket{3}$ is a linear combination of $\ket{2}, \ket{4}, \Omega\ket{4}$. Then we may write $\ket{3} = \alpha \Omega \ket{2}+\beta \ket{4} + \gamma \Omega\ket{4}$ for some real numbers $\alpha, \beta, \gamma$.
Taking the inner product with $\bra{4}$ yields $0=\braket{4|3} = \beta \braket{4|4}$, so $\beta=0$.
This implies $a_{23} = \braket{2|3} = \beta\braket{2|4}=0$ as asserted. \hfill $\square$

We have shown that it suffices to only consider states whose covariance matrix is symmetric, symplectic, $a_{33}=a_{44}$, and satisfies $a_{ij}=0$ if $i+j$ is odd.
The only covariance matrices which satisfy these conditions are those of TMSV states with additional signal mode squeezing.
\begin{align}\label{Vvar}
V&=S \begin{pmatrix} \cosh 2\zeta & \sigma_z \sinh 2\zeta \\ \sigma_z \sinh2\zeta & \cosh 2\zeta \end{pmatrix}  S^T ,\\
S&= \text{diag} (e^{\zeta_s}, e^{-\zeta_s} , 1, 1).
\end{align}
Therefore, we only need to optimize over three parameters, the TMSV parameter $\zeta$, the single-mode squeezing parameter $\zeta_s$, and the displacement $r=\sqrt{x_s^2 + p_s^2}$.

We now return to maximizing the monotone metric when the input state to QI is a Gaussian state with covariance matrix as in Eq. (\ref{Vvar}) and first-order moments $(x_s,p_s,0,0)^T$.
Using our general result Eq. (\ref{GaussianMetric}), we need to maximize
\begin{align} 
g_{\rho_0} (d\rho, d\rho) &=  2  \frac{ \sinh^2 2\zeta \cosh^2 \zeta_s}{(\nu_1-1)(\nu_2-1)}  f\left(\frac{\nu_1+1}{\nu_1-1} \frac{\nu_2+1}{\nu_2-1} \right)^{-1} \nonumber \\
& + 2\frac{ \sinh^2 2\zeta \sinh^2 \zeta_s}{(\nu_1+1)(\nu_2-1)}  f\left(\frac{\nu_1-1}{\nu_1+1} \frac{\nu_2+1}{\nu_2-1} \right)^{-1} \nonumber \\
&+  \frac{2 r^2}{\nu_1-1} f \left( \frac{\nu_1+1}{\nu_1-1} \right)^{-1},
\end{align}
with $\nu_1 = 1+2N_B$ and $\nu_2 = \cosh 2\zeta$, under the constraint $r^2+\cosh 2\zeta \cosh 2\zeta_s = 1+2N_S$.
For convenience, we define $\zeta_b:=\frac{1}{2}\cosh^{-1} \nu_1$.
A general integral representation of operator monotone functions gives
\begin{equation}
\frac{1}{x} f\left( \frac{y}{x} \right)^{-1} = \int_0^1 \left[\frac{1}{xs+y}+\frac{1}{x+ys} \right] d\sigma(s)
\end{equation}
where $d\sigma(s)$ is a positive measure on $[0,1]$ \cite{AL99}.
This means that we need to maximize
\begin{widetext}
\begin{align}\begin{split}\label{IntRep}
g_{\rho_0} (d\rho, d\rho) &=  \int_0^1 \left[\frac{2\sinh^2 \zeta \cosh^2 \zeta_s}{s \cosh^2 \zeta_b + \sinh^2 \zeta_b \tanh^2 \zeta}+\frac{2\sinh^2 \zeta \cosh^2 \zeta_s }{ \cosh^2 \zeta_b + s \sinh^2 \zeta_b \tanh^2 \zeta} \right.\\
& \hspace{15mm} +\frac{2\sinh^2 \zeta \sinh^2 \zeta_s }{s \cosh^2 \zeta_b \tanh^2 \zeta + \sinh^2 \zeta_b }+\frac{2\sinh^2 \zeta \sinh^2 \zeta_s }{ \cosh^2 \zeta_b \tanh^2 \zeta + s \sinh^2 \zeta_b} \\
& \hspace{15mm} + \left.\frac{ r^2}{s \cosh^2 \zeta_b + \sinh^2 \zeta_b}+\frac{ r^2}{ \cosh^2 \zeta_b + s \sinh^2 \zeta_b} \right] \sigma(s) ds.
\end{split}\end{align}
\end{widetext}
We will show that the term in the square bracket is maximized for TMSV states regardless of $s$, hence TMSV states maximize any monotone metric of the two output states of QI.

Define the function $F$ as follows to use Lagrange multipliers:
\begin{align} \begin{split}
F(\zeta, \zeta_s, & r, \lambda) = \text{(Square bracket of Eq. (\ref{IntRep}))} \\
&+ \lambda(r^2+\cosh 2\zeta \cosh 2\zeta_s - 1-2N_S ).
\end{split} \end{align}
We have the following three cases.

Case 1. If $r\neq 0$, the $\partial F/\partial r =0$ equation determines $\lambda$.
Substituting $\lambda$ into the $\partial F / \partial \zeta_s =0$ equation gives $\zeta_s =0$ which further yields $\zeta =0$ from the $\partial F / \partial \zeta =0$ equation.
Hence coherent states are critical points.

Case 2. If $r=0$, $\zeta_s\neq 0$, the  $\partial F / \partial \zeta_s =0$ equation determines $\lambda$.
Substituting $\lambda$ into the $\partial F / \partial \zeta =0$ equation gives $\zeta =0$.
Hence single-mode squeezed vacuum states are critical points.

Case 3. If $r=0$, $\zeta_s =0$, we get TMSV states.

Therefore, coherent states, single-mode squeezed vacuum states, and TMSV states are all of the critical points. The values of each are
\begin{widetext}
\begin{align}
F(0,0,\sqrt{2N_S},\lambda) &= \frac{ 2N_S}{s \cosh^2 \zeta_b + \sinh^2 \zeta_b}+\frac{ 2N_S}{ \cosh^2 \zeta_b + s \sinh^2 \zeta_b} \hspace{15.55mm}\text{(for coherent states)},\\
F(0,\sinh^{-1} \sqrt{N_S},0,\lambda) &= 0 \hspace{36.55mm} \text{(for single-mode squeezed vacuum states states)},\\
F(\sinh^{-1} \sqrt{N_S},0,0,\lambda)&= \frac{ 2N_S}{s \cosh^2 \zeta_b + \frac{N_S}{1+N_S} \sinh^2 \zeta_b}+\frac{ 2N_S}{ \cosh^2 \zeta_b + s \frac{N_S}{1+N_S} \sinh^2 \zeta_b} \; \text{(for TMSV states)}.
\end{align}
\end{widetext}
It is evident that TMSV states maximize $F$ among the three values.
Also, the difference between coherent states and TMSV states decreases as $N_S\rightarrow \infty$, which gives evidence that the quantum advantage is prominent in the low signal mean photon number regime.

\section{Optimal Single Mode Input for QI} \label{App:SMQI}
In Ref. \cite{MB21}, the authors claim that coherent states are the optimal probe in single-mode QI.
We fill in a gap in their argument in this section.
Starting with a pure state $\ket{\psi} = \sum_n c_n \ket{n}$, the authors show that maximizing $\gamma_{\textit{col}}$ is equivalent to maximizing $\gamma:= \sum_n c_{n+1}c_n \sqrt{n+1}$ under the constraints $\sum_n c_n^2 =1$ and $\sum_n n c_n^2 = N_S$ where we may assume that $c_n$ are real.
The Lagrange multiplier method gives the following equation:
\begin{equation}
c_{n+1} \sqrt{n+1} +c_{n-1} \sqrt{n} +2c_n (\mu_1 + n\mu_2) =0, \label{EQ:LM}
\end{equation}
where $\mu_1$ and $\mu_2$ are Lagrange multipliers.
As coherent states satisfy Eq. (\ref{EQ:LM}), the authors of \cite{MB21} conclude that coherent states are the optimal probe for single-mode QI.

We find all solutions to Eq. (\ref{EQ:LM}) and indeed show that coherent states are the optimal probe for single-mode QI.
First observe that Eq. (\ref{EQ:LM}) can be written concisely as
\begin{equation}
(2\mu_2 \hat{a}^\dagger \hat{a} +\hat{a} + \hat{a}^\dagger +2\mu_1) \ket{\psi} =0.
\end{equation}
Applying the displacement operator $\hat{D} := \text{Exp} (\frac{\hat{a}^\dagger - \hat{a}}{2\mu_2})$ leads to
\begin{equation}
\left(2\mu_2 \hat{a}^\dagger \hat{a} +2\mu_1 - \frac{1}{2\mu_2}  \right) \hat{D} \ket{\psi} = 0,
\end{equation}
which means that $\hat{D} \ket{\psi}$ is a number state and $\ket{\psi}$ is a displaced number state.
Therefore, solutions to Eq. (\ref{EQ:LM}) are the displaced number states
\begin{align}
\ket{\psi} = \hat{D}^\dagger \ket{n},\; \mu_2 = -\frac{1}{2\sqrt{N_S-n}}
\end{align}
with $n<N_S$.
We compute $\gamma$ as
\begin{equation}
\gamma = \braket{\psi|\hat{a}|\psi} = - \frac{1}{2\mu_2} = \sqrt{N_S -n}
\end{equation}
which is obviously maximized for $n=0$, i.e., $\ket{\psi}$ is a coherent state.

\end{document}